\documentclass{svproc}

\usepackage{graphicx}	
\usepackage{amsmath}	
\usepackage{amssymb}	

\begin{document}
%
%

\title{The origin of galaxy scaling laws in LCDM}
\titlerunning{Galaxy scaling laws in LCDM}  
%
\author{Julio F. Navarro\inst{1}}
\authorrunning{Julio F. Navarro} 
%
\tocauthor{Julio F. Navarro}
\institute{Physics and Astronomy, University of
  Victoria, Victoria, BC, V8P 5C2,  Canada. \\
\email{jfn@uvic.ca}}
\maketitle              
\begin{abstract}
  It has long been recognized that tight relations link the mass,
  size, and characteristic velocity of galaxies. These scaling laws
  reflect the way in which baryons populate, cool, and settle at the
  center of their host dark matter halos; the angular momentum they
  retain in the assembly process; as well as the radial distribution
  and mass scalings of the dark matter halos. There has been steady
  progress in our understanding of these processes in recent years,
  mainly as sophisticated N-body and hydrodynamical
  simulation techniques have enabled the numerical realization of
  galaxy models of ever increasing complexity, realism, and
  appeal. These simulations have now clarified the origin of these  galaxy scaling laws in a universe dominated by cold dark matter:
  these relations arise from the tight (but highly non-linear)
  relations between (i) galaxy mass and halo mass, (ii) galaxy size
  and halo characteristic radius; and (iii) from the self-similar mass
  nature of cold dark matter halo mass profiles. The excellent agreement between simulated and observed galaxy scaling laws is a resounding success for the LCDM cosmogony on the highly non-linear scales of individual galaxies.
\end{abstract}
\section{Introduction}
The current paradigm for structure formation envisions a Universe whose matter component is dominated by cold dark matter (CDM) and whose recently accelerating expansion reflects the negative pressure of a mysterious form of ``dark energy'' that resembles Einstein's cosmological constant (Lambda, or ``L'', for short).  The nature of the dark matter and the source of dark energy constitute our era's premier challenges to our understanding of the physical Universe. Unraveling the nature of dark matter, in particular, is
widely seen as the most promising way to extend the well-established
Standard Model of Particle Physics, and is one of the most cherished
goals of contemporary Theoretical Physics.

Detailed modeling of the cosmic microwave background
(CMB) and of large-scale galaxy clustering have led to a few widely
accepted conclusions: dark matter is almost certainly non-baryonic (or
behaved as such at the time of primordial nucleosynthesis); it
dominates roughly 5:1 over normal, baryonic matter, and clusters on a
wide range of scales, from galaxy super-clusters to dwarf
galaxies \cite{Planck2018}. 

Averaged over large scales, dark matter is distributed
throughout the Universe in a web-like structure that matches closely
that expected to arise from gravitational amplification of Gaussian
density fluctuations. The fluctuation amplitude dependence on scale is
also well constrained, and is broadly consistent with that expected from nearly scale-free perturbations in a collisionless fluid with small or negligible thermal velocities; i.e., ``cold dark matter'' (L+CDM, or ``LCDM'', for short). These successes imply that, at least in the quasi-linear regime probed by scales larger than about a
small galaxy group, any successful model of dark matter must be or
behave like CDM.

On smaller scales there are no observational probes of the linear power
spectrum, and, therefore, the clues rely on the clustering of
dark matter inferred from observations in the highly non-linear regime of individual
galaxies. Since the galaxy baryonic component often plays a substantial role on
these scales, the evidence is indirect, the predictions rely heavily on numerical simulations, and the interpretation can be ambiguous or
inconclusive. Indeed, a number of ``challenges'' to LCDM have been
identified on dwarf galaxy scales that, although probably not lethal to LCDM,
have attracted keen attention from advocates of modifications to LCDM
or even to our standard model of gravity \cite{Bullock2017}.

The purpose of this contribution is to add to this discussion by
assessing the health of the LCDM paradigm on the non-linear scales of
individual galaxies. I focus on observations on the scale of the ``$L_*$'' galaxies where  the large majority of stars in the Universe reside \cite{Li2009,Baldry2012}. In particular, I describe the origin, in LCDM, of
the Tully-Fisher relation (TFR) that links the rotation speed of a
disc galaxy with its stellar/baryonic mass, as this is a sensitive test of the predicted non-linear clustering of cold dark matter.

\section{The Tully-Fisher relation}
\label{SecTFR}
The Tully–Fisher relation \cite{Tully1977} is a particularly useful galaxy scaling relation because it links, with tight scatter, distance-dependent and distance-independent quantities. Properly calibrated, the TFR can therefore be used as a secondary distance indicator to map out the cosmic flows in the Local Universe and to measure Hubble's constant. 

The observed TFR has long challenged direct numerical simulations of disc galaxy formation in LCDM. Indeed, early work produced galaxies so massive and compact that their rotation curves were steeply declining and, at given galaxy mass, peaked at much higher velocities than observed \cite{Navarro2000,Scannapieco2012}.

\begin{figure}[t]
\includegraphics[width=\textwidth]{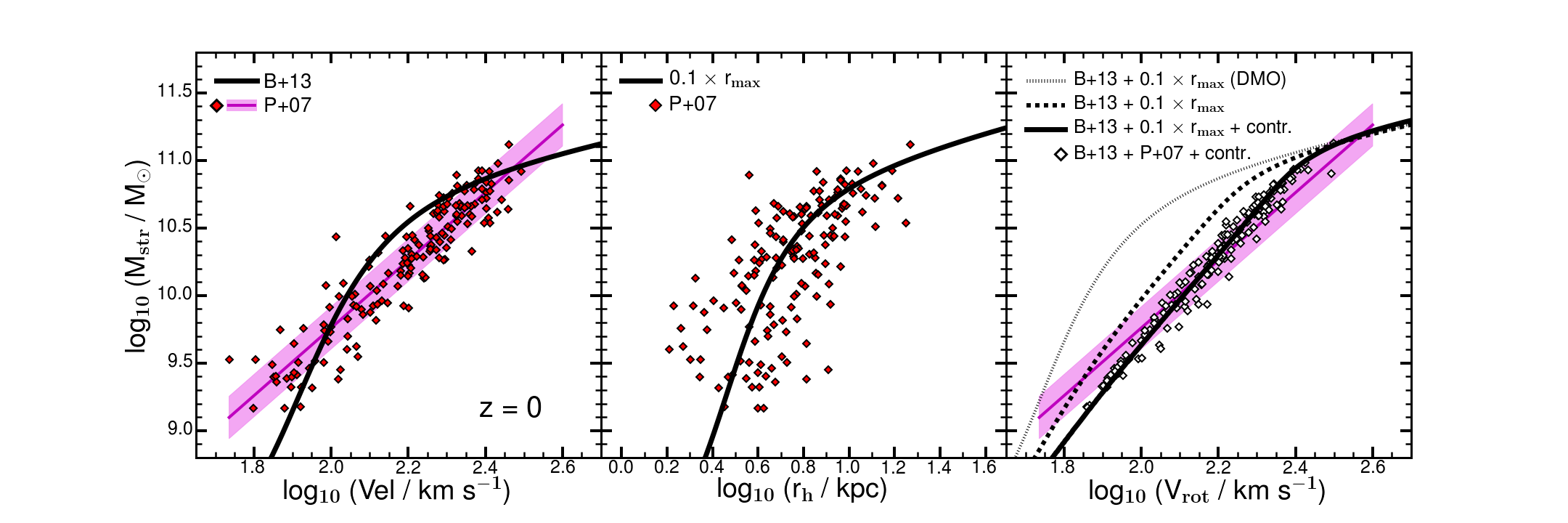}
\caption{Galaxy stellar mass, $M_{\rm str}$, as a function of various
  parameters. {\it Left:} The solid black curve shows the
  abundance-matching prediction of \cite{Behroozi2013}, as a
  function of halo virial velocity, $V_{200}$. Symbols correspond to
  the data of \cite{Pizagno2007}, converted to stellar masses
  using a constant I-band mass-to-light ratio of $1.2$ and shown as a function of disk rotation speed,
  $V_{\rm rot}$. Color-shaded band indicates the mean slope and
  $1$-$\sigma$ scatter. {\it Middle:} Symbols show half-light radii of
  galaxies in the P+07 sample. Thick solid line indicates a multiple
  of $r_{\rm max}$, the characteristic radius where NFW halo circular
  velocities peak. Halo masses are as in the \cite{Behroozi2013} model of
  the left panel.  {\it Right:} Tully-Fisher relation. The color band
  is the same as in the left-hand panel. The dotted curve indicates
  the dark halo circular velocity at $r_h=0.1\, r_{\rm max}$, assuming
  NFW profiles and neglecting the contribution of the disk. The dashed
  line includes the gravitational contribution of the disk, keeping
  the halo unchanged. Finally the thick solid line (and symbols)
  include the disk contribution {\it and} assume that halos contract
  adiabatically. This figure taken verbatim from \cite{Ferrero2017}.}
\label{FigTFR}
\end{figure}

\begin{figure}[t]
\begin{center}
\includegraphics[width=80mm]{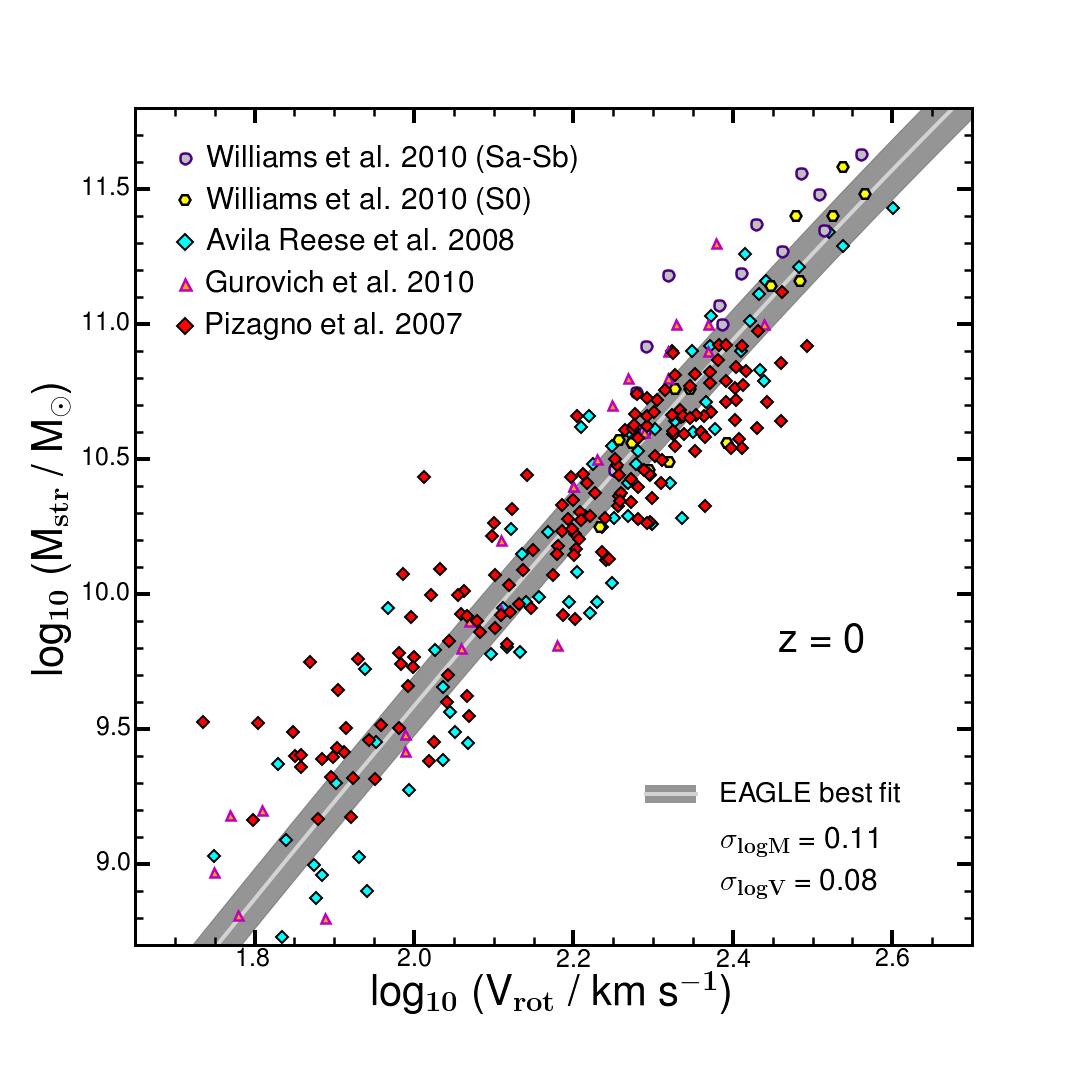}
\caption{Tully-Fisher relation for EAGLE galaxies (grey band) compared
  with individual spirals taken from five recent TF compilations. 
  The simulated relation is in excellent agreement with the
  observational data. The scatter is even smaller than in observed
  samples, even though the simulated relation includes {\it all}
  galaxies and not only disks. This figure taken verbatim from \cite{Ferrero2017}.}
\label{FigTFR2}
\end{center}
\end{figure}

The rotation speed of a disc depends on its baryonic mass and size
(which set the contribution of the luminous component to the circular
speed), as well as on the dark mass contained within the disk
radius. The latter depend on the radial mass profile of dark matter halos, which is self-similar and well described in LCDM by the ``NFW profile'' \cite{Navarro1997}. The contribution of the dark matter to the circular velocity of a disc galaxy, then, depends only on the relation between galaxy mass and halo mass. 

This relation is in turn fully constrained by the galaxy stellar mass function through the ``abundance matching'' (AM) approximation \cite{Behroozi2013}. The AM ranks galaxies by mass and assigns them to halos ranked in similar fashion, preserving the ranked order. Satisfying this approximation appears to be a {\it sine qua non} condition for any cosmological simulation that attempts to reconcile the LCDM halo mass function with the galaxy stellar mass function \cite{Schaye2015}. This implies that there is no extra freedom in LCDM to ``tune'' the Tully-Fisher relation, making the TFR a useful probe of the clustering of dark matter distribution in the highly-non-linear scales of
individual galaxies.

One feature of the AM approximation is that it predicts a complex relation between galaxy mass and halo mass. We show this in the left panel of Fig.~\ref{FigTFR}, where the solid black line indicates the AM-derived halo virial\footnote{The virial mass of a halo, $M_{200}$, is conventionally defined as that enclosed within a radius, $r_{200}$, where the mean density is 200 times $\rho_{\rm crit}=3H_0^2/8\pi G$, the critical density for closure. Virial quantities are measured at that radius and are listed with a ``200'' subscript.} velocity (which is equivalent to halo virial mass; see $X$-axis) as a function of galaxy stellar mass ($Y$-axis). Disc rotation speeds for a sample of galaxies (the observed TFR) are shown by the symbols in the same panel. The observed relation clearly differs in shape and normalization from the AM relation between galaxy stellar mass and halo virial velocity, $V_{200}$.

However, disc rotation velocities are measured at the
half-light radius, $r_{\rm h}$, of the galaxy, and not at the virial radius. Half-light radii for the same galaxy sample are shown in the middle panel of Fig.~\ref{FigTFR}. Measuring dark matter circular velocities at $r_{\rm h}$ leads to smaller values than $V_{200}$ (grey curve on the left of the right-hand panel of Fig.~\ref{FigTFR}) because, at  $r_{\rm h}$, the NFW halo circular velocity profile is still rising. This is even more at odds with the observed velocities. Adding the contribution of the baryonic disc, however, yields higher velocities, as indicated by the thick dotted line in the same panel. Finally, accounting for the response ("adiabatic contraction'') of the halo to the assembly of the galaxy yields the thick solid line. This very crude model reproduces quite well the zero point and scatter of the TFR, as may be judged by the excellent agreement between the thick solid line and the symbols, which represent  the model resluts when applied to the individual galaxies of the sample. The slope is slightly off from the observed relation, but this is a shortcoming of the approximate model adopted to represent the halo contraction. Indeed, a cosmological hydrodynamical simulation where galaxy disc masses roughly agree with AM and disc sizes agree with observation results in a TFR in excellent agreement with the observed relation \cite{Ferrero2017}, as shown by Fig.~\ref{FigTFR2}.

\section{Outlook}

We stress that the success of LCDM in accounting for the TFR is not simply a result of parameter tuning. Once the cosmological parameters are specified, if galaxies are assigned to halos so as to reproduce the galaxy stellar mass function and the galaxy mass-size relation is roughly in agreement with observation, then the resulting mass-velocity scaling for disc galaxies matches the observed TFR strikingly well. In other words, CDM halos add ``just the right amount'' of dark matter to the luminous regions of galaxies so as to reproduce the TFR. This is a non-trivial result that should be rightfully regarded as a true success of the LCDM cosmogony.

Key to this success is the self-similar "NFW" mass profile of LCDM. This profile implies that galaxies form in regions where the circular velocity of the halo is steadily rising, and where dark matter contributes a sizable, but not dominant, fraction of the mass enclosed within the half-light radius. The NFW profile shape is responsible for the rather small dispersion of the TFR:  galaxies of different mass and size that populate halos of a given mass spread {\it along} the TFR, thus minimizing the scatter. We conclude that the TFR is a sensitive and telling test of the predicted clustering of CDM on the highly non-linear scales corresponding to the half-light radii of disc galaxies. LCDM passes this test with flying colors. 

Other galaxy scaling laws can also be used to test the predicted structure of LCDM halos. One example is the mass discrepancy-acceleration relation (MDAR), which links the spatial distribution of baryons with the speed/acceleration at which they orbit in galaxy discs \cite{McGaugh2016}. This has also been examined in LCDM by a number of authors, who converge to conclude that the MDAR is just a reflection of the self-similar nature of cold dark matter halos and of the physical scales introduced by the galaxy formation process \cite{Navarro2017}.

We have not examined here some of the small-scale challenges to LCDM highlighted in other work, and expertly reviewed by \cite{Bullock2017}. These include the "missing satellites" and "rotation curve diversity" problems, the "too-big-to-fail" puzzle, the "missing dark matter galaxies", and the "cusp-core" controversy. We have addressed several of them in recent contributions, including \cite{Sawala2016,Fattahi2016,Oman2015,Oman2016}, and have argued that all of them admit plausible resolutions in LCDM. The LCDM paradigm thus seems in excellent health, and news of its demise will, in the opinion of this author, prove exaggerated.

\bigskip

I acknowledge the hospitality of the KITP at UC Santa Barbara.  This research was supported in part by the National Science Foundation under Grant No. NSF PHY-1748958.

\bibliography{simons-symp-refs}
\clearpage
\bibliographystyle{spphys.bst}

\end{document}